# Predicting Vulnerability in Large Codebases With Deep Code Representation

Anshul Tanwar, Krishna Sundaresan, Parmesh Ashwath,
Prasanna Ganesan, Sathish Kumar Chandrasekaran, Sriram Ravi

*Cisco Systems*

***Abstract*– Currently, while software engineers write code for various modules, quite often, various types of errors - coding, logic, semantic, and others (most of which are not caught by compilation and other tools) get introduced. Some of these bugs might be found in the later stage of testing, and many times it is reported by customers on production code. Companies have to spend many resources, both money and time in finding and fixing the bugs which would have been avoided if coding was done right. Also, concealed flaws in software can lead to security vulnerabilities that potentially allow attackers to compromise systems and applications. Interestingly, same or similar issues/bugs, which were fixed in the past (although in different modules), tend to get introduced in production code again.**

***We developed a novel AI-based system which uses the deep representation of Abstract Syntax Tree (AST) created from the source code and also the active feedback loop to identify and alert the potential bugs that could be caused at the time of development itself i.e. as the developer is writing new code (logic and/or function). This tool integrated with IDE as a plugin would work in the background, point out existing similar functions/code-segments and any associated bugs in those functions. The tool would enable the developer to incorporate suggestions right at the time of development, rather than waiting for UT/QA/customer to raise a defect.***

***We assessed our tool on both open-source code and also on Cisco codebase for C and C++ programing language. Our results confirm that deep representation of source code and the active feedback loop is an assuring approach for predicting security and other vulnerabilities present in the code.***

- *Anshul Tanwar - PRINCIPAL ENGINEER E-mail: atanwar@cisco.com.*
- *Krishna Sundaresan - VP ENGINEERING E-mail: ksundar@cisco.com*
- *Parmesh Ashwath - SOFTWARE ENGINEER E-mail: parmesh20120@gmail.com*
- *Prasanna Ganesan - DIRECTOR ENGINEERING E-mail: prasgane@cisco.com*
- *Sathish Kumar C - TECHINCAL LEADER E-mail: sathicha@cisco.com*
- *Sriram Ravi – MANAGER ENGINEERING E-mail: srravi@cisco.com*

## I. INTRODUCTION

Developers are at the core of every organization. They write tons of code and sadly, many bugs as well. And many of those bugs get shipped to production, and that's unavoidable. The code always has visible or hidden flaws, which if gets missed by coder, can only be caught by subsequent activities, such as code-reviews and testing. Using Machine learning and AI techniques, we can improve the developer experience while coding and avoid potential bugs that might occur in future even before the next stages. Billions of lines of source code that have been written contain implicit knowledge about how to write good code, code that is easy to comprehend and to debug. We can leverage the wealth of available data complemented with state-of-art machine learning models to develop the enterprise-level solutions to have a high standard of coding & potentially less bugs and thus helping enterprise to save resources.

There are traditional methods like static and dynamic analyzers to help developers improve the code quality. Both static and dynamic analyzers are rule-based tools and thus are limited to hand-engineered rules. These tools further introduce too many false positives making a few critical findings buried in the sea of warnings and also, these don't take application-specific field/production errors into account for the future analysis. Even, on identifying a possible issue, no further insights on its impact nor the knowledge of potential fix are provided for the developer. We developed novel AI based system to overcome this.

Feature extraction from source code is an active area of research being worked on by many organization and open source community. There exist many different robust techniques to represent the source code as a feature set and fed into the machine learning models for achieving a different task. In our work, we propose a novel method of representing a function code using a distributed vector of fixed length similar to work done by Uri Alon et al. [1] with a subtle difference of considering ranked path context which will be discussed in detail later in the modelling section.

The code vectors obtained also follows all the convention that the word embeddings [2] does. The vectors of two similar functions will be closer to each other in the N-Dimensional space. And also, the combination and analogies property hold true on the vectors obtained. We will be leveraging this property later in our workflow to identify the similar function in the database given a code vector of a function of interest. This vector also forms the basis for our vulnerability prediction task.

The code vectors obtained are used as a feature vector to train a model for identifying the vulnerabilities present in the source code. We employed a unique approach involving a feedback loop for active learning to incorporate the knowledge of identifying the potential application logic errors.

For vulnerabilities detection task, the model was trained using both open-source and proprietary code base and bug repository. The details of which will be highlighted in the data section below.

We also studied different approaches to deployment and integration strategies for our solution to make it work at an enterprise level. We developed an AI plugin which will be available to the existing Integrated Development Environment (IDE) and assist the developers while writing the code with similar functions and also the potential vulnerabilities present in the code if any. The details of the working prototype will be shared in later sections.

The rest of the paper will be organized in the following sections. In the Related Work section, here we focus on the existing methods which are addressing a similar problem. In the Data Source section, here we talk about the different sources used for our analysis and model building activity. In the Data Preparation and Feature Extraction section, we will explain how the code vectors were obtained from the dataset and also, we will talk about the data labelling task and approach carried out for our proprietary codebase. The Model section will then cover the different classification models trained and tested for the above-prepared data. And finally, we will talk about the results we have achieved and also the deployed strategies we followed to move our model to production.

## II. RELATED WORK

Distributed representations of words (word2vec) [2], sentences, paragraphs, and documents (doc2vec) [3] are considered a milestone in the field of neural network and natural language processing (NLP). These papers lead to many groundbreaking modeling techniques later. Similar approach was extended to get the distributed representation for the code block using an Attention-based network in [1]. For learning the code representation, the method names are taken as label and the method label predication is taken as primary task. Having a good method name not only helps in modeling activity but also it helps in general for project maintenance [4,5]

There is also another line of work where there is significant research done to represent source code as a token stream. However, in general, deep code representation is found to be more effective than considering source code as just linear sequence of tokens [6].

For the Vulnerability detection task there are many tools available in market like Clang, FlawFinder, CppCheck, and Coverity [7,10]. As mentioned before these tools are rule driven.

Beyond the traditional tools, several machine learning driven techniques are also employed for vulnerability detection task. Simple bag-of-words technique was used to extract feature and train an SVM for classification work was done in [11]. The work in [12] expanded on this work by including n-grams in the feature vectors used with the SVM classifier. In [13] the potential of deep learning was explored for program analysis by encoding the nodes of the abstract syntax tree representations of source code and training a tree-based convolutional neural network (CNN) for classification problems. Work in [14] considered a similar feature extraction technique but employed the recurrent neural network (RNN).

In [15] work has been done on using deep learning to learn features directly from source code in a large natural codebase to detect a variety of vulnerabilities using CNN based network for feature extraction and then neural network for classification problem.

In the work done in [16], they have considered the code embeddings which represents the semantic structure of the code block alone for bug prediction. Here they uses the code embeddings obtained by code2vec[1] and run a binary classification for off-by-one errors using synthetic data created by mutation of comparator operators.

To our knowledge, there is no work carried out to perform the vulnerability detection using code embeddings learned from the source code by incorporating the functionality of a module as well and making it work taking enterprise bug repository into account.

## III. DATA SOURCE

Given the complexity and heterogeneity of programs, a vast number of training examples are needed to train machine learning models that can adequately learn the distributed code representation or embeddings. We mined many open-source code repositories present in GitHub [17] to extract function level code and also Cisco codebase to learn the code embeddings. As learning the code embeddings task takes the function name as the label for primary prediction task, it requires no other data labelling allowing us to consider any well written open-source code. We tried to include different flavored code including OS codebases like Linux, Debian packages [18], application logic code, network driver code, API protocol code and many others. A total of around **8.9 million functions** were used to train the model used to extract the code embeddings.

For Vulnerability detection task, we need a labeled dataset to train a classifier model. We initially started with SATE IV Juliet Test Suite [19]. While the SATE IV dataset provides labeled examples of many types of vulnerabilities, it is made up of synthetic code snippets that do not sufficiently cover the space of natural code to provide an appropriate training set alone. Also, we found it to be so unnatural/repetitive to be almost dangerous for model training. It's easy to overfit on the data and not realize it since the train/test splits will have a lot of similar functions. So, we had to drop the usage of Juliet dataset. For initial model training we then mainly leveraged Draper VDISC Dataset - Vulnerability Detection in Source Code [20]. The dataset consists of the source code of **1.27 million functions** mined from open source software, labelled by static analysis for potential vulnerabilities. The tools used for labelling includes Clang, Cppcheck, and Flawfinder. After labeling, significant effort was done from the authors of the dataset to clean up the duplicates and remove the false labelling. The labels considered for analysis are CWE119, CWE120, CWE469, CWE476 and other CWE's are grouped into CWEOthers category.

After preliminary modelling to enable the feedback loop, we considered our proprietary bugs and security vulnerability database. The functions which could have potentially caused the historical bugs are taken from the database and fed to the model for its prediction and insights. Using this custom database is optional, however its usage and manual validation of model results enables the fine-tuned model to identify the potential application logic errors which are critical for an enterprise wide adoption.

## IV. DATA PREPARATION & FEATURE EXTRACTION

We believe that have a novel approach of data cleaning, data preparation and feature extraction from the source code.

The source code (from Cisco codebase) is converted to an Abstract Syntax Tree (AST) representation. Using AST, each method is represented with the set of encoded path context. We have a unique approach here to compute the number of path context representation for a given function. Initially, we take all path context representation from AST and then we run a ranking model to eliminate some of them which are occurring commonly across the functions and also the path contexts that are found in very few functions so that we avoid overfitting of the model by increasing the nodes in the first layer of the model to account the input dimensionality later. The minimum number of the occurrence of a path context is one of the hyperparameters of our model. This method is found to work very effectively when compared to considering all the path context.

Let P represents the set of path contexts in an AST. Each path context (p) will be of format ni-pij-nj. Where ni and nj represents the encoded node values and pij represents the encoded path values.

Node encodings are the numerical representation of each node in an AST. Similarly, path encodings represent the numerical representation of each path in an AST. These numerical encodings help us to convert the text to a number which can be then be used to feed the model training as input.

Each path contexts obtained will then be filtered to remove the path contexts which occurs very frequently as they don't help in uniquely distinguish the AST or the function of interest and also those which occurs very rarely so that we will not have higher dimensionality in the encodings.

*Filtered path contexts ($P_F$) =* $\forall\, p \in P\,, if\ min \leq count(p)\ \geq max$

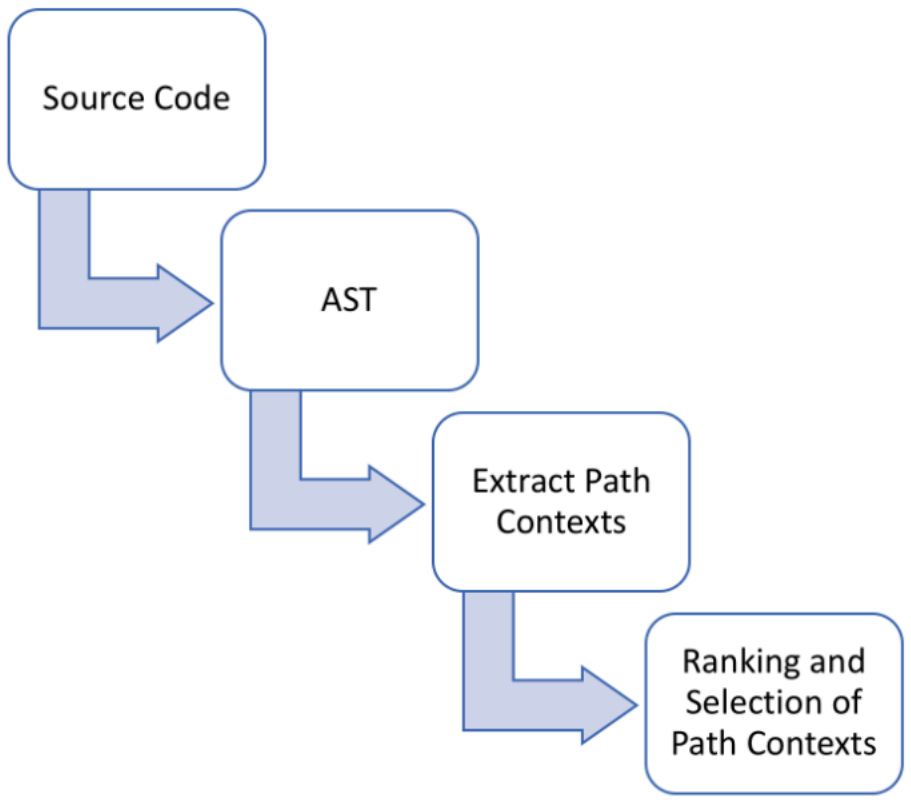

The next task here is to get a single vector representation also referred to as code embeddings or code vector for a function from the set of path context. We train a path context - Attention-based model to learn the code vectors. The code vectors would be the weighted average of the path embedding concatenated with the weighted average of the node embedding.
Node embeddings and path embeddings are learned from the encodings during the model training
The Attention weights are learned from training the model with the task of method name prediction as a primary task with SoftMax as the output layer. The originality of this approach is in the fact of using both path embeddings and node embeddings to obtain path context and selecting them using an Attention model.

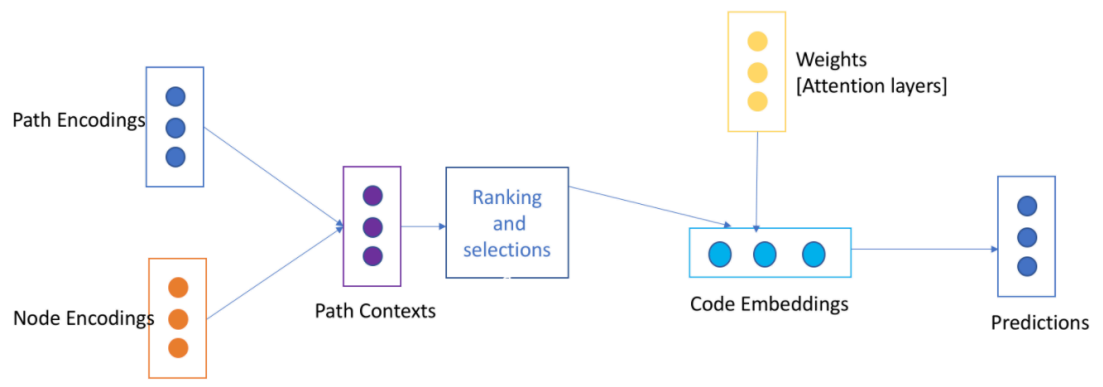

For each p in PF, during the model training, we learn the path context embeddings ci,

$$c_i = embeddings(n_i, p_{ij}, n_j)$$
$$ci = [node_{embeddings(n_i)} \quad path_{embeddings(p_{ij})} \quad node_{embeddings(n_j)}]$$

Now to obtain the single embedding representation for a given function, we take the weighted average of all ci. The weights are learned from the attention layers

$$code\ embedding = \sum_{i=1}^{n} \alpha_i * c_i$$

$$where\ \alpha_i = \frac{exp\ (c_i^T . a)}{\sum_{j=1}^{n} c_j^T . a}$$

Where a denotes the global attention vector which is initialized randomly and learned simultaneously with the network

With the above two steps, we get a vectorized representation or the code embeddings of each function present in Cisco codebase. The code embeddings are dumped to a file at function granularity. The file follows the standard format as word embeddings, each line follows a format of the function name followed by its corresponding code vector. The code vector is of length 384 in our use case, but this is a configurable parameter. As mentioned above, we consider both node and path encodings to arrive at the final vector representation. The node encodings and path encodings are each of length 128.

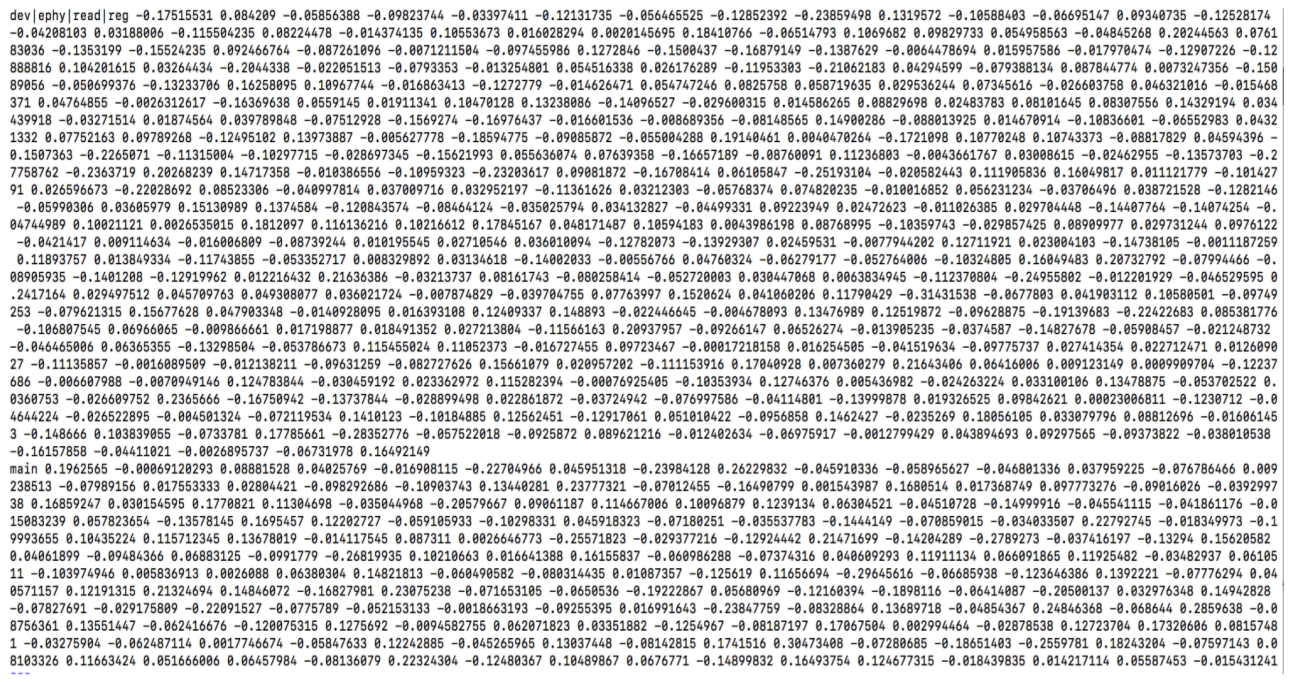

The vectors obtained and shown above also follows all the convention that the word embeddings do. The vectors of two similar functions will be closer to each other in the N-Dimensional space. And also, the combination and analogies property hold good on the vectors obtained. This vector forms the basis for our vulnerability prediction and similar function detection task.

Using the trained model, we extract the vectors for each function present in the Draper VDISC Dataset and finally we will have the below labeled dataset:
X → code vector of a function
Y → 5 class of CWE's (1- if present,0- if not)

The code embeddings obtained are at the function granularity and it mainly deals with semantic structure of the code as it was formed using AST's. In addition to this we will also consider another set of embeddings to represent the functionals aspect of the block of code which is referred to as composite code embeddings later in this paper.

## V. Modeling Workflow

Once we have the code embeddings ready for each function in our training set, we follow below approach which involves an active feedback loop to incorporate the application logic error findings. Below are the three phases involved in this workflow.

### A. Training & Prediction Phase:

- Using the above trained code embeddings model, we extract the code vectors for every function present inside Draper VDISC dataset. This acts as our initial labelled dataset consisting of code vectors as independent variables and the marked CWE's labels as dependent variables.
- For training the vectors, we consider two sets of 3-layer Neural network. One trained with just the vanilla code embeddings and another model trained with composite code embeddings. Here the composite embeddings include the complete functionality of the module and help to identify the errors spread across multiple functions. And the first one help us to identify the function semantic and other errors within the block of code. We then use a simple logistic regression model to combine the results of both of the model so that appropriate weightage is given to both vanilla and composite code embeddings.

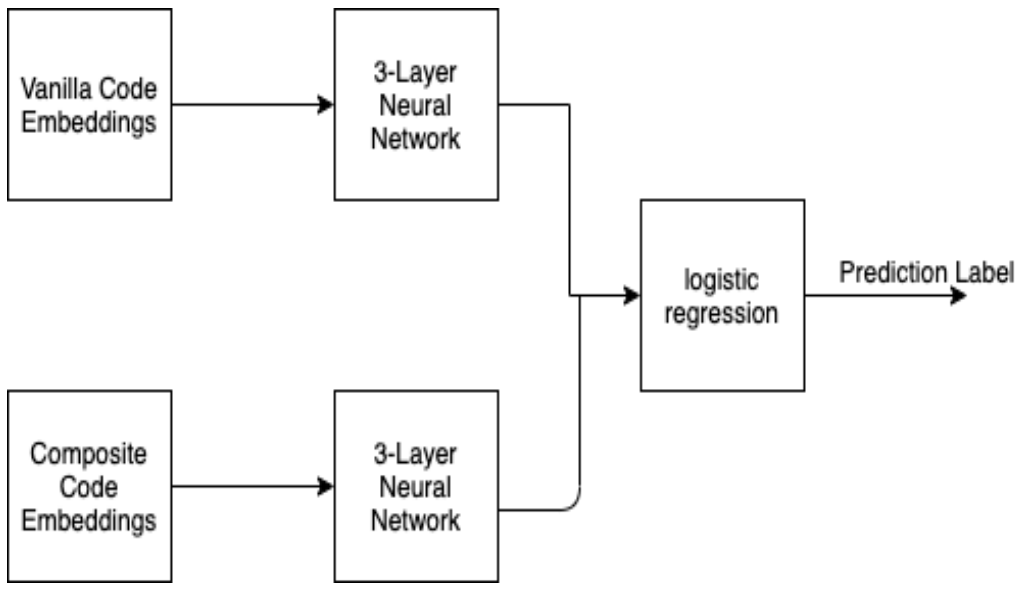

- To evaluate the performance of the model and for further fine-tuning, historical Cisco Bug Data is considered.
- For every Bug, we analyze the Diff Enclosures and take the previous version of the functions present, which might have caused the bug.
- For each function obtained, we run the model to predict its probability for having any of the CWE or bugs in general

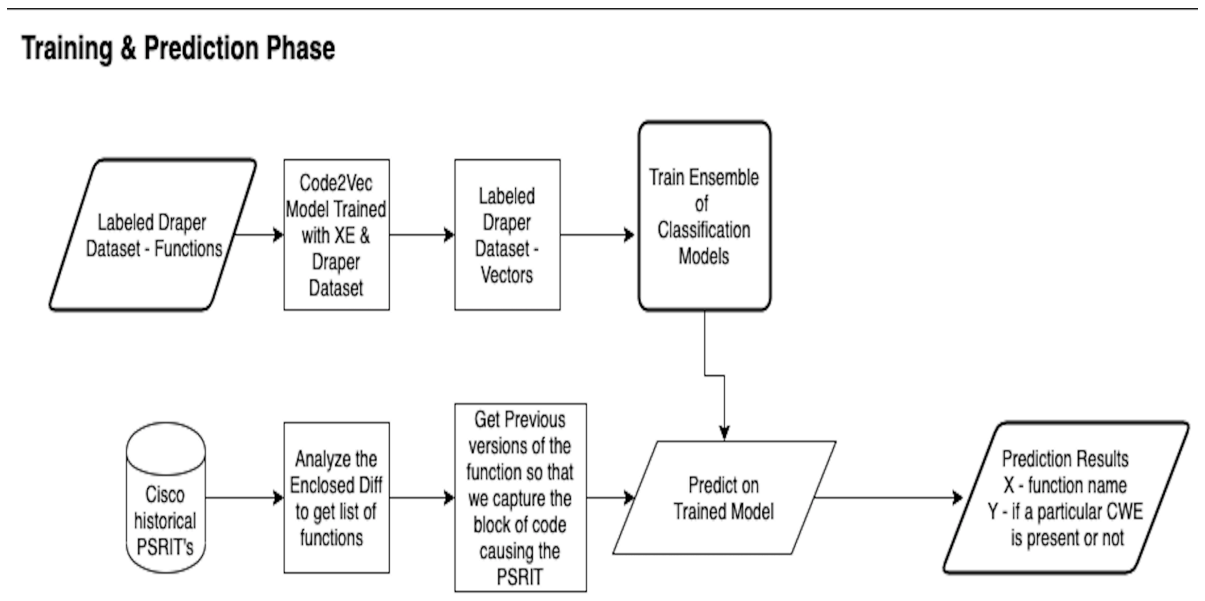

### B. Manual Validation to Fine-Tuning the Model

- The results obtained by the models are validated using 2 approaches:
  - Perform Manual Validation on each predicted result. During this analysis, developers will also mark an additional label indicating if the code contains issues like application logic error [other than CWE's]
  - We run a set of SA tools [Flawfinder, CPPCheck, Coverity, Clang] and see if the predicted CWE's is identified by anyone of them.
- Using the above feedback data, we fine-tune the model parameters and finalize the model that will be used for each CWE prediction.
- Finally, we will re-train the above model using the Cisco PSRIT validated data [from the above step]. This time we include application logic error as another dependent variable

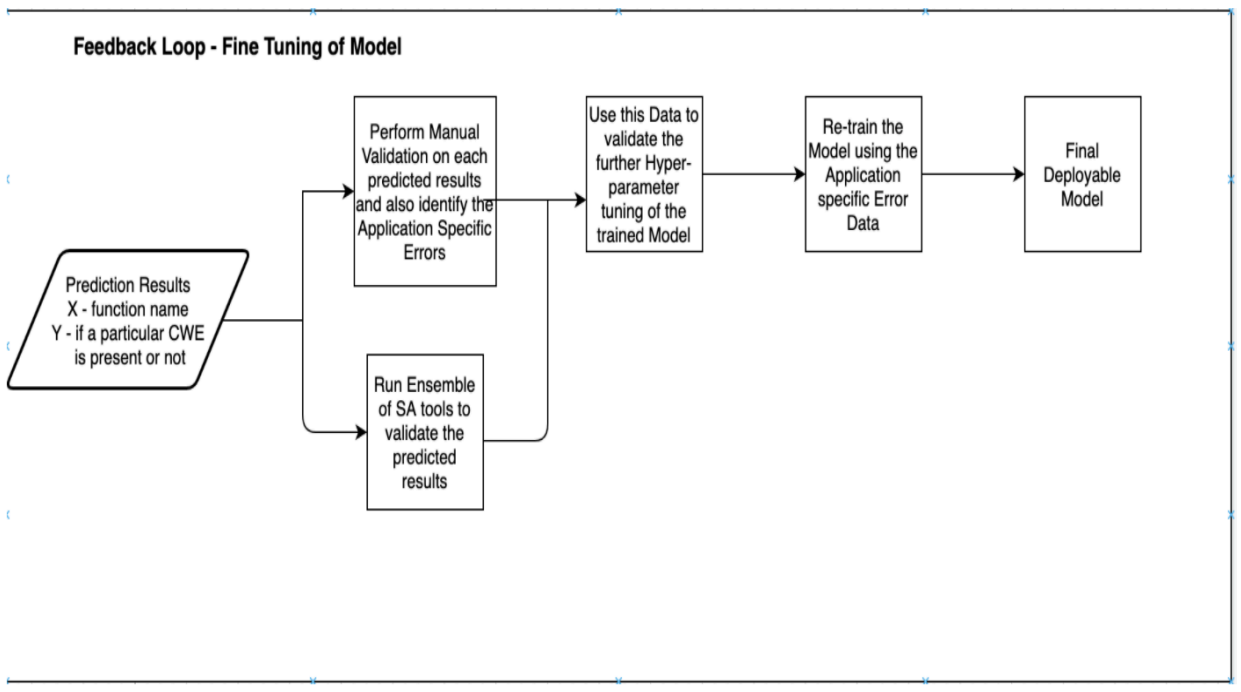

This feedback loop is what makes our approach eliminate many false positives and also include the application logic errors in its findings.

### C. Production Deployment of the Model

We follow 2 strategies for model deployment in Production

- **Batch Mode**: Run the model prediction for every function in Enterprise codebase, starting with fewer component and share the results and insights with the developer for Bug cleanup.

- **Real-Time**: Integrate the model prediction as a plugin in IDE, so whenever a developer writes a new block of code We provide the details as the chance of the code being buggy, also suggest probable fixes wherever possible. The Fix is suggested by looking for the similarity of the given function to the historical Bugs causing function and then taking the fix that was done to close the bug. The similarity check happens in the N-Dimensional space using the code2vec representation

## VI. Deployment & Predictions

After the model is trained using the approach detailed in the above section, the next task is to carry out the predictions for **similar function identification** and **bug prediction.**

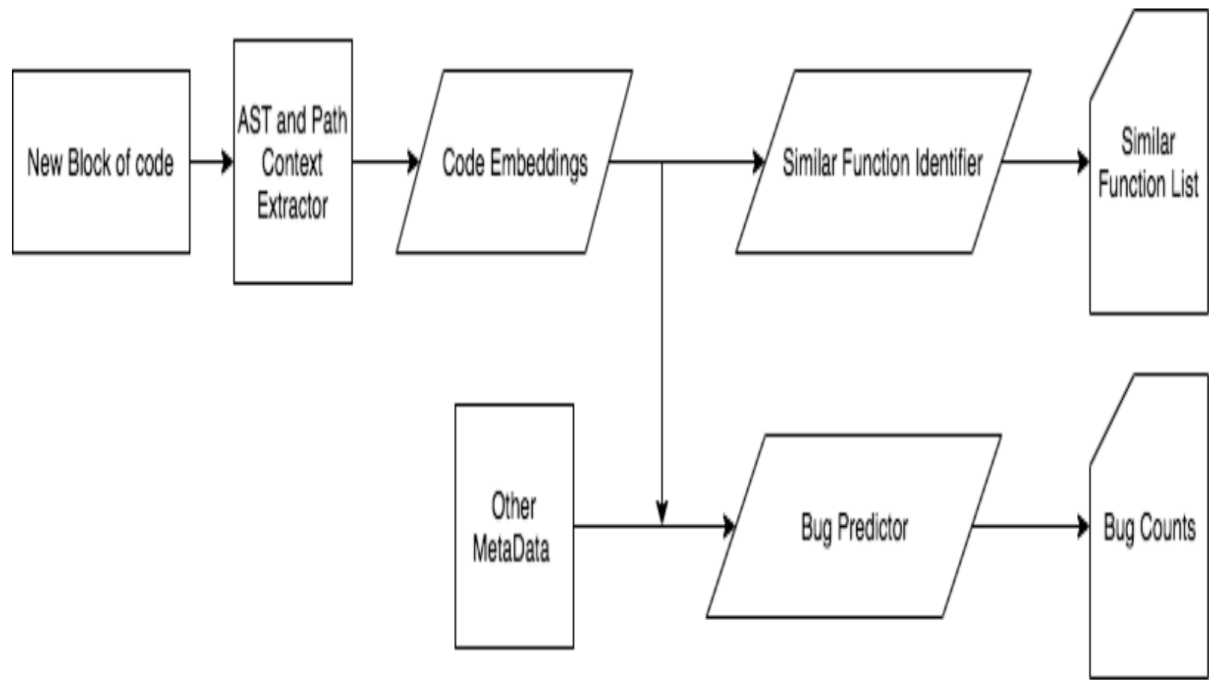

We have developed two models here to predict the following:

1. The number of potential bugs in a newly developed function: The code embeddings learnt above is one of the main features for our model of bug prediction. The other features that were of interest are the static analysis score, function coverage, hotspot score and also the developer information. With all the above feature we train an ensemble of models which includes Fully connected Neural network, Regression models to predict the number of bugs.
2. For the other task of finding similar functions, we again use an ensemble technique of KNN and clustering methods. The code embeddings in the main feature used for this.

## VII. Feedback Loop for Active Learning

The developer has an option to provide direct feedback to the model about the potential bugs that are being prevented by the tool and ensures that our system results do not stagnate. This also has a significant advantage in that this data used to train new versions of the model is of the same real-world distribution that the customer cares about predicting over.

To incorporate the feedback, we will modify the code embeddings that the model estimated for a given function based on the user feedback using the below strategy

• For positive feedback/vote from the user, we will take the code embeddings of a new function that was being developed and move it closer to the function tagged with the predicted bug by a certain distance in N-dimensional space. The distance moved is proportional to the logarithm of the number of positive votes.

• For negative feedback/vote from the user, we will take the code embeddings of a new function that was being developed and move it farther to the function tagged with the predicted bug by a certain distance in N-dimensional space. The distance moved is proportional to the logarithm of the number of negative votes.

In addition to this, we will also retrain the model from scratch taking the current embeddings as the initial weights for every function in our database that includes the embedding for the recently added functions also so that we will maintain the model with the up-to-date codebase. Frequency of training can be decided by the domain experts and currently, model retraining happens every month.

## VIII. Results

We also carried out the vulnerability task in two iterations, in first one we did not consider composite code embeddings nor the additional logistic regression and below are results we obtained

For similarity function identification task, we have arrived at a threshold on the distance to be less than **0.4** for two functions to be similar and this thresholding technique is giving us the accuracy of **95%.**

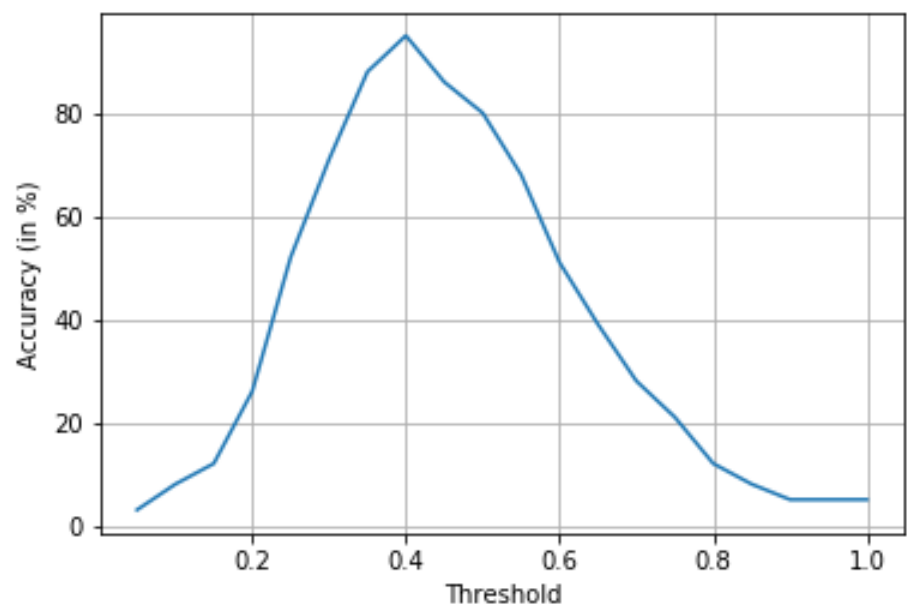

*Figure 1: Accuracy vs Threshold Variation*

For Bug Prediction task, we used historical raised bugs and see if our model predicts them beforehand on the functions associated with them. Here we had an accuracy of around **70%**. Also, we got the precision value of **0.74** and a recall of **0.77.**

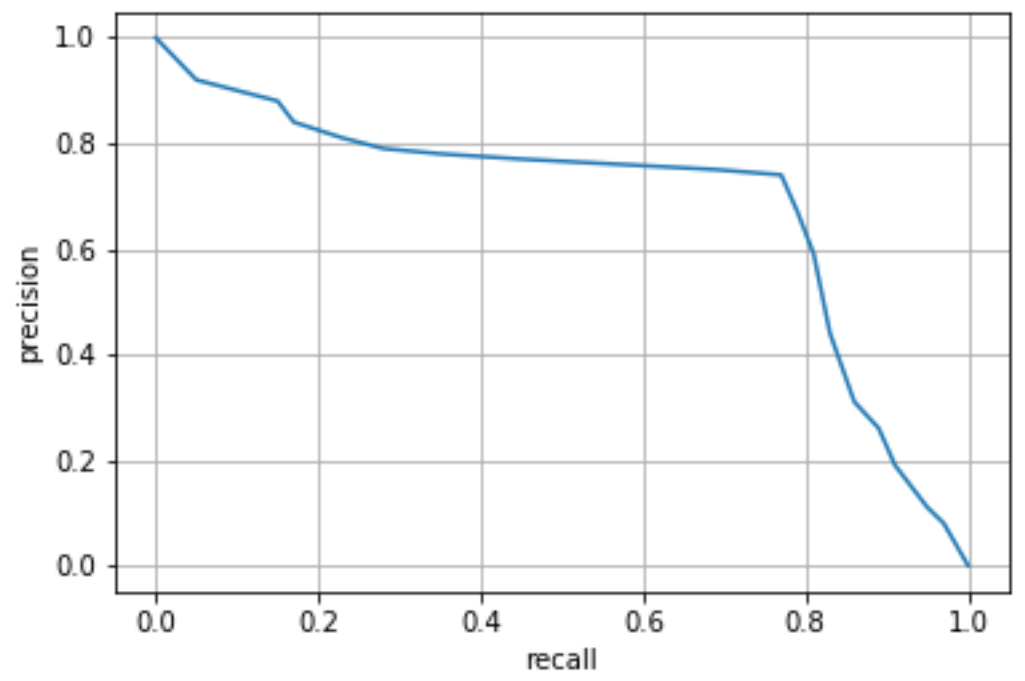

*Figure 2: Precision vs Recall Curve for Bug Prediction*

In the second iteration by including composite code embeddings, we saw a significant boost in the performance of the same bug prediction task described above, we achieved an accuracy of around **78%**. The new precision value was **0.81** and a recall of **0.82.**

## IX. Conclusion

We presented a new technique to train accurate and robust neural models of code. Our work addresses key challenges inherent to the domain of code. The Deep code representation technique presented here can be used as the base feature extraction technique on the code and the extracted code embeddings can form as the basis to many machine learning tasks.

To address the second challenge of Vulnerability detection, we proposed a new workflow involving a novel feedback loop approach to determine the potential bugs and application logic errors present in the code. The new workflow also takes the composite code embeddings into account for bug prediction and our work highlighted the improvements achieved by doing so. We also presented an enterprise level integration for the ML models done here.

## Acknowledgments

The authors thank Cisco Systems for allowing us to carry out this research work and also, we would like to thank the domain experts and developers who helped us to manually validate the model initial results that helped us to improve our modeling techniques and fine-tuning of model.